\documentclass{article}%
\usepackage{amsmath}
\usepackage{amsfonts}
\usepackage{amssymb}
\usepackage{graphicx}
\usepackage{multirow}
\usepackage{fleqn}

\title{A marvelous contribution from Michel H\'{e}non to globular cluster's study :
the isochrone cluster}
\author{J\'{e}r\^{o}me Perez\\
Applied Mathematics Laboratory, ENSTA-ParisTech, Palaiseau, France}
\begin{document}

\maketitle{}
Globular clusters are cornerstones in the study of the dynamics of
gravitationaly interacting systems of particles.

As a matter of fact, they are made of thousands of stars -- at minimum -- and,
in the contrary of galaxies, they do not contain gas or other dissipative
components. Hence, excepting the fact that stars are not punctual, they
corresponds exactly to their mathematical modelisation.

The observation of globular clusters reveals that they are characterized by a
spherical distribution of their mass.

By consequence, their volumic mass density is a radial function $\rho
=\rho\left(  r\right)  $. This density corresponds to the marginal
distribution in position space of the total distribution function in the whole
phase space: this is a mean field description. In this model the
gravitational potential $\psi$ of the cluster is solution of the Poisson
equation $\Delta\psi=4\pi G\rho$, it is then radial too: $\psi=\psi\left(
r\right)  $. The $r$ variable represents the distance between the center of
mass of the system and a test star evolving in the mean field potential of the
cluster. This test star of mass $m$ experiences a force given by
$\boldsymbol{F}=-m\boldsymbol{\nabla}\psi$, which is already radial. \ In this
context, it is very well known that its trajectory is contained in a plane. In
this plane, the orbit is determined by the total energy $E$ and the squared
angular momentum $L^{2}$ of the star. These two quantities are conserved
during the motion.

The description of the whole dynamics of a globular cluster is then possible
when its gravitational potential $\psi$ is given. Three possibilities exists
to do this:

\begin{itemize}
\item Extract $\rho\left(  r\right)  $ and $\psi\left(  r\right)  $ from
observational data.\ It is a rough and empirical manner, but it is necessary
to fix ideas.

\item Compute $\rho\left(  r\right)  $ and $\psi\left(  r\right)  $ from
numerical simulations.\ But the parameter's space of the result is wide for
this kind of experiments.

\item Propose fundamental physical arguments in order to select a model
through all theoretical possibilities.
\end{itemize}

The two first ways are very used and a plenty of references give
refined to rough descriptions of globular clusters since their formation until
the end of their actual or numerical evolution (see \cite{HH2003}). The last
way is by far the least used. Michel H\'{e}non's paper \cite{amasi} about
isochrone cluster, published in french in "Annales d'Astrophysiques" is very
representative of this way to do which count no more than a ten of reference
in more than two centuries of globular cluster modelisation! In this paper,
we will follow Michel H\'{e}non's paper in a first part to analyze in a second
part its influence under gravitational dynamicists community, giving finally a
modern reading of the result.

\section{The isochrone model}

As we mention before, the motion of a given star in a spherical
globular cluster is contained in a plane. In this plane the two parameters of
the orbit are its energy $E$ and its squared angular momentum $L^{2}$. Both
these two parameters contribute to the definition of the gravitational
potential of the cluster $\psi\left(  r\right)  $ and to the computation of
the distance $r$ between the star and the center of mass of the cluster. This
contribution is resumed in the definition of the energy of the star%
\begin{equation}
E=\frac{m}{2}\left(  \frac{dr}{dt}\right)  ^{2}+\frac{L^{2}}{2mr^{2}}%
+m\psi\left(  r\right)  =\mathrm{cste}\label{defE}%
\end{equation}

Imposing that the total mass of the system\footnote{The total mass
$M\ $of a spherical self-gravitating system is given by the integral of the
density over the whole space. In spherical coordinates, using the Poisson
equation, this gives
\par%
\[
M=\frac{1}{G}\left[  \lim_{r\rightarrow\infty}\left(  r^{2}\frac{d\psi}%
{dr}\right)  -\lim_{r\rightarrow0}\left(  r^{2}\frac{d\psi}{dr}\right)
\right]
\]
} is finite, the effective potential
\[
\psi_{e}(r)=\frac{L^{2}}{2mr^{2}}+m\psi\left(  r\right)
\]
is such that $\lim_{r\rightarrow0}\psi_{e}(r)=+\infty$.

At the edge of the system, up to an additive constant, the potential is on the
form $\psi\left(  r\right)  _{r\rightarrow+\infty}\sim-GM/r$, where $M$ the
total mass of the cluster. Hence, $\lim_{r\rightarrow+\infty}\psi_{e}%
(r)=0^{-}$.

\begin{figure}[ptb]
\begin{center}
\includegraphics[scale=1]{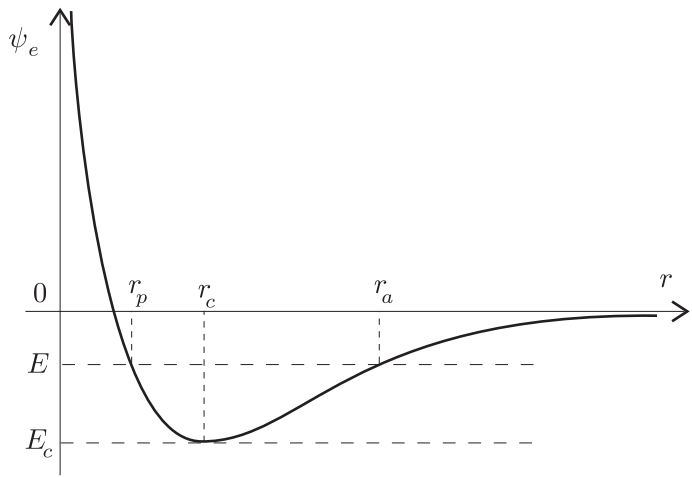}
\end{center}
\caption{Effective potential of a spherical cluster}%
\label{poteff}%
\end{figure}

This effective potential is also the place where $\dfrac{dr}{dt}=0$, its
behavior is represented on figure \ref{poteff}. 

 When the considered star belongs to the cluster its total energy $E$
is negative. The extreme value $E=E_{c}<0$ corresponds to an orbit for which
$r=r_{c}=\mathrm{cste}$, it is then circular. For each value of $E\in\left]
E_{c},0\right[  $, the distance is such that $r\in\left[  r_{p},r_{a}\right]
$.\ The extreme values of the radius are the periastron for $r_{p}$ and the
apoastron for $r_{a}$. The time for the transfer from $r_{p}$ to $r_{a}$ is
given by
\begin{equation}
\frac{\tau}{2}=\int_{t_{a}}^{t_{p}}dt=\int_{r_{a}}^{r_{p}}\frac{dt}%
{dr}dr={\displaystyle\int_{r_{a}}^{r_{p}}}\frac{dr}{\ \sqrt{2\left[  \dfrac
{E}{m}-\psi\left(  r\right)  \right]  -\dfrac{L^{2}}{m^{2}r^{2}}}%
}\ \ \ \label{perad}%
\end{equation}
where we have used the energy to get $\dfrac{dr}{dt}$. The problem being
symmetrical, if this integral is finite, we have $r_{a}=r\left(  t_{a}\right)
=$ $r\left(  t_{a}+\tau\right)  $. The expression (\ref{defE}) of the energy
is in fact an ordinary differential equation, fulfilled by both $r\left(
t\right)  $ and $r\left(  t+\tau\right)  $. These two functions are then equal
when $t=t_{a}$, the Cauchy criterion -- which is supposed valid for this
problem -- then tell us that they are equal at each time : the distance
between the center of mass of the cluster and the star is then a $\tau
-$periodic function. The inspection of the relation $\left(  \ref{perad}%
\right)  $ shows that, if it exists, the radial period is such that $\tau
=\tau\left(  E,L^{2}\right)  $. The mass of the star is another parameter of
the period which is not under interest in this study. The radial period is
computable at least in two fundamental cases:

\begin{itemize}
\item The Kepler potential $\psi\left(  r\right)  =-GM/r$, where $M$ is the
total mass of the cluster. \ It is the well-known two body problem for which
Kepler's third law gives $\tau=\frac{\pi GM}{\sqrt{2}}\left(  -E\right)
^{-3/2}$. It is also the limit potential viewed by a star of the cluster
always evolving in far from center regions.

\item The harmonic potential $\psi\left(  r\right)  =\frac{2}{3}\pi G\rho
r^{2}$, where $\rho$ is the constant density of a homogeneous cluster, for
which $\tau=\sqrt{\frac{3\pi}{4G\rho}}$.\ Using the virial theorem, one can
show that $\rho$ depends only of the energy of the considered star and the
total mass of this homogeneous system. The central regions of a globular
clusters are usually considered as homogenous ones, the gravitational
potential is then harmonic at the center of such systems. A centrally confined
star could then have a radial period depending only of its energy $E.$
\end{itemize}

 Both centrally or far from center confined orbits are then
characterized by radial periods which do not depends on the squared angular
momentum. In his 1958 paper, Michel H\'{e}non conjectures that this property
could propagates to all orbits of all stars of a globular system. \ This so
called "isochrone" property, could be a fundamental physical property allowing
to determine the other properties of globular clusters.

Thus, Michel H\'{e}non proposes to find the most general potential $\psi
_{i}\left(  r\right)  $ such that%
\[
\tau=2{\displaystyle\int_{r_{a}}^{r_{p}}}\frac{dr}{\ \sqrt{2\left[  \dfrac
{E}{m}-\psi_{i}\left(  r\right)  \right]  -\dfrac{L^{2}}{m^{2}r^{2}}}}%
=\tau\left(  E\right)
\]

\begin{figure}[ptb]
\begin{center}
\includegraphics[scale=1]{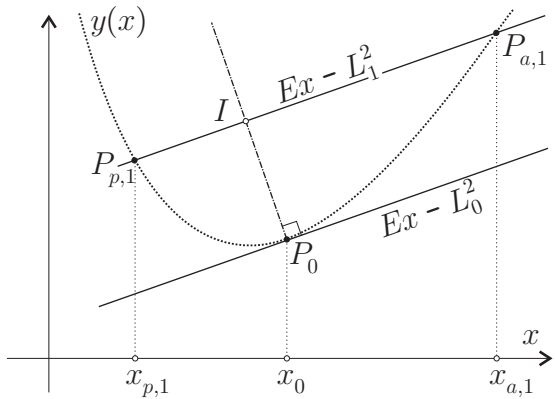}
\caption{Illustration from Michel H\'{e}non paper about the characteristic points and $y\left(x\right)$ function.}%
\label{fig1}%
\end{center}
\end{figure}

To do this he doesn't use the effective potential but introduces new
variables%
\[
x=2r^{2}\text{ \ and }y\left(  x\right)  =x\psi_{i}\left(  x\right)  \text{.}%
\]
Fixing $m=1$, the radial period then writes
\[
\tau=\int_{x_{a}}^{x_{p}}\frac{dx}{\sqrt{Ex-L^{2}-y\left(  x\right)  }}%
\]
$x_{p}$ and $x_{a}$ being the values of $x$ to the periastron and apoastron,
roots of the equation $y\left(  x\right)  =Ex-L^{2}$. Due to the finite mass
of the system, $y\left(  x\right)  $ must have infinite branches. In order to
identify $x_{p}$ and $x_{a}$, Michel H\'{e}non then propose a graphical study
of the problem. This representation is an alternative of the effective
potential theory, it is full of sense and forms a new way for the study of
such systems.\ This graphical study is illustrated in the figure \ref{fig1}.

By an explicit calculus of the radial period in terms of $x$ and $y$, Michel
H\'{e}non shows that $\tau$ depends only on the energy $E$ if and only if
\[
P_{0}I\propto\left(  P_{p,1}P_{a\,1}\right)  ^{2}%
\]
This condition is fulfilled if and only if $y\left(  x\right)  $ is a parabola.

The physical parameters of the problem are
\[
\psi_{0}=\lim_{r\rightarrow0}\psi\left(  r\right)  \text{, }\psi_{\infty}%
=\lim_{r\rightarrow+\infty}\psi\left(  r\right)  \text{ and the total mass
}M\text{.}%
\]
Three cases can be considered:
\begin{itemize}
\item If $\psi_{0}\rightarrow-\infty$, it is always possible to choose a
reference for the potential such that $\psi_{\infty}=0$.\ The equation for
$y\left(  x\right)  $ is then
\[
y\left(  x\right)  =-GM\sqrt{2x}\ \ \implies\ \ \ \psi_{i}\left(  r\right)
=-\frac{GM}{r}%
\]
It is the keplerian potential associated to a central mass.

\item If $\psi_{0}\rightarrow-\infty$, it is always possible to choose a
reference for the potential such that $\psi_{0}=0$. The equation for $y\left(
x\right)  $ is now%
\[
y\left(  x\right)  =\frac{1}{2}Kx^{2}\ \ \implies\ \ \text{ }\psi_{i}\left(
r\right)  =Kr^{2}%
\]
It is the harmonic potential.

\item If both  $\psi_{0}$ and $\psi_{\infty}$ are finite, choosing a reference
for the potential such that $\psi_{\infty}=0$, the equation for $y\left(
x\right)  $ becomes%
\[
y^{2}+\frac{G^{2}M^{2}}{2\psi_{0}}y-2G^{2}M^{2}x=0\implies\ \ \text{ }\psi
_{i}\left(  r\right)  =\frac{2\psi_{0}}{1+\sqrt{1+\frac{r^{2}}{b^{2}}}%
}\ \ \text{with\ }b=-\frac{GM}{2\psi_{0}}>0
\]
this is the general form introduced by Michel H\'{e}non for his potential, the
isochrone potential.
\end{itemize}

The paper ends by a comparison between this new potential and the
globular clusters data observation of this epoch. Michel H\'{e}non finds a
good agreement and conjectures a physical explanation: a resonance between
orbits during the formation process of globular clusters could vanishes the
$L^{2}$ dependence of the radial period. The history will forget progressively
the isochrone model mainly because more and more observations will reveal that
the potential density pair is trickier than a simple stationary model. In 1968
the \emph{empirical} King model, with 3 free parameter, became the reference.
Does the isochrone model, with only the parameter\ $b$, could be forgotten? It
should be a big error...

\section{A modern lecture of the isochrone model}

Isochrone cluster are characterized by a central structure of
constant density, namely a core, surrounded by a halo. The size of the core is
given by the parameter $b$ of the model. The radial period of the isochrone
model can be explicitly computed, it is given by the relation
\[
\tau=\frac{2\pi GM}{\left(  -2E\right)  ^{3/2}}%
\]
Using Poisson equation one can give an explicit formula for the mass density
\[
\rho\left(  r\right)  =\frac{M}{4\pi}\frac{3ab\left(  a+b\right)  -br^{2}%
}{a^{3}\left(  a+b\right)  ^{3}}\text{ \ \ with }a^{2}=r^{2}-b^{2}%
\]
When $r\gg b$, i.e. in the halo, the mass density is given by a power law
$\rho\left(  r\right)  \propto r^{-4}$.\ This property correspond to young
globular clusters, i.e. characterized by a large two body encounters
relaxation time ($T_{c}$) in comparison of the Hubble time.

In a numerical way, the equilibrium state with a core and a $-4\ $halo slope
structure corresponds to the post collapse state of an initial
homogeneous sphere. These kind of systems are generally called "H\'{e}non Spheres" 
since the pioneering numerical experiments of Michel H\'{e}non in the sixties
\cite{sphenon}. They could correspond to one of the globular clusters
formation process. After this formation process which takes a few dynamical
times ($T_{d}$), such a system evolves under collisional effects under very
more longer times, order of them is given by the two body encounters
relaxation time ($T_{c}\simeq\frac{N}{\ln N}T_{d}$, where $N$ is the number of
stars in the system). During this slow evolution the extension of the core
shrinks progressively and the slope of the halo passes from $-4$ to $-2$. At the
end of this evolution, the system become unstable and the core collapses under
the pressure of the halo.

On the century of globular clusters orbiting in our galaxy, 80\% have
a core-halo structure with slope in the interval $\left[  -4,-2\right]  $ and
20\% are said core-collapsed. The fine mechanisms of this evolution begins to
be well understood, but, the question of the initial equilibrium state at the
beginning of this evolution is always an open problem. In this context, the
Michel H\'{e}non's isochrone idea seems to be one of the possible physical
explanations. 

Very few research are done in order to verify Michel H\'{e}non's conjecture
about the resonant mechanism as the origin of the isochrone model. A
confirmation of the existence of such a mechanism could be an important and
elegant fact in the comprehension of globular clusters, in the direct line of
Michel H\'{e}non works.

\end{document}